          [2001/04/25 1.1 (PWD)]
\documentclass[a4paper,twocolumn]{esapub} 
\usepackage{natbib}
\usepackage{epsfig}
\usepackage{txfonts}

\title{\huge Bering -- The first deep space mission to map asteroidal diversity, origin and transportation}
\author{Anja C.\,Andersen}
\affil{NORDITA, Blegdamsvej 17, 2100 Copenhagen, Denmark, E-mail: anja@nordita.dk}
\author{Ren\'e Michelsen}
\affil{Astro.\ Obs., NBIfAFG, Juliane Maries Vej 30, 2100 Copenhagen Denmark, E-mail: rene@astro.ku.dk}
\author{Henning Haack}
\affil{Geological Museum, {\O}ster Voldgade 5-7, 1350 Copenhagen K, Denmark, E-mail: hh@savik.geomus.ku.dk}
\author{John L.\,J{\o}rgensen}
\affil{{\O}rsted*DTU, MIS, Building 327, Tech.\ Uni.\ of Denmark, 2800 Lyngby, Denmark, E-mail: jlj@oersted.dtu.dk}

\begin{document}

\maketitle

\begin{abstract} Asteroids are remnants of the material
from which the Solar System formed. Fragments of asteroids, in the form of meteorites, include samples of  the first solid matter to form in our Solar System 4.5 mia years ago. Spectroscopic studies of asteroids show that they, like the meteorites, range from very primitive objects to highly evolved small Earth-like planets that differentiated into core mantle and crust. The asteroid belt displays systematic variations in abundance of asteroid types from the more evolved types in the inner belt to the more primitive objects in the outer reaches of the belt thus bridging the gap between the inner evolved apart of the Solar System and the outer primitive part of the Solar System. High-speed collisions between asteroids
are gradually resulting in their break-up. The size
distribution of kilometer-sized asteroids implies that the
presently un-detected population of sub-kilometer asteroids
far outnumber the known larger objects. Sub-kilometer
asteroids are expected to provide unique insight
into the evolution of the asteroid belt and to the meteorite-asteroid
connection. We propose a space mission to
detect and characterize sub-kilometer asteroids between
Jupiter and Venus. The mission is named “Bering” after
the famous navigator and explorer Vitus Bering. A
key feature of the mission is an advanced payload package,
providing full on board autonomy of both object detection
and tracking, which is required in order to study
fast moving objects in deep space. The autonomy has the
added advantage of reducing the cost of running the mission
to a minimum, thus enabling science to focus on the
main objectives.
\end{abstract}

 \vspace*{-0.3 cm}

\section{Introduction}

 \vspace*{-0.4 cm}
Our present understanding of asteroids and their orbits is almost entirely 
based on surveys of main-belt asteroids  with diameters larger
than 10 km. Ground based 
telescopes cannot detect smaller objects except within the immediate vicinity 
of Earth and no spacecraft has, so far, detected any previously 
unknown asteroids. Despite the fact that several spacecrafts to date, 
statistically, must have passed by smaller asteroids, the technology 
employed in these vessels has not held the capability of 
detecting these objects. Therefore such encounters have gone unnoticed by. Recent 
development in the autonomy of space-borne image- and computer-technology 
has changed this, so that it is now possible to detect, classify and 
observe during an encounter with a small asteroid.

The sub-kilometer objects between Jupiter and Venus, in particular the 
Near-Earth Asteroids (NEAs), are expected to fill the
gap between the meteorites that we have studied in very great detail in the
laboratory and their large parent asteroids in the main belt that may be
studied with Earth-based telescopes. The meteorites have been knocked off their
parent asteroids through impacts. These impacts delivered fragments in a large
range of sizes.
Streams of small asteroids are connected to parent asteroids via dynamical
mechanisms responsible for the transfer of material to the inner 
Solar System [1].

\begin{figure}[]
\centering
\epsfig{file=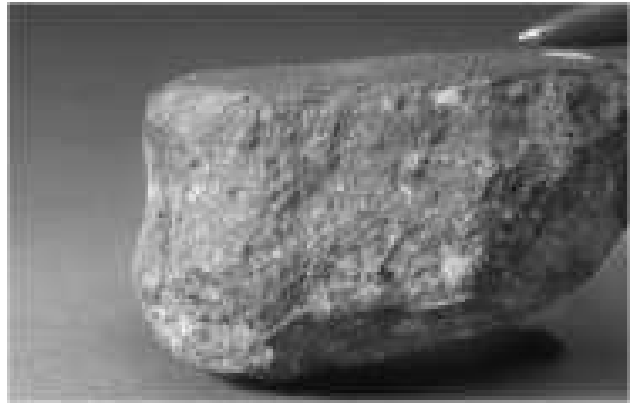,width=1.0\linewidth}
\caption{A fragment of the carbonaceous chondrite Allende that fell in Mexico
in 1969. The meteorite is composed of dark fine grained dust, mm-sized spherical
inclusions (chondrules) and white inclusions know as calcium-aluminum-rich inclusions (CAIs). The CAIs formed 4567 My ago and are the oldest known solids 
formed in the Solar System. Carbonaceous chondrites probably originate from C-type asteroids which are common in the outer main-belt.}
\label{fig:meteorite}
\end{figure}

Meteorites are highly diverse geological samples of asteroids, the Moon and Mars. 
They range from very primitive samples of the first solids to form in the Solar 
System (Fig.\ 1) to highly evolved samples of differentiated planetary objects. The latter include 
iron meteorites from asteroid metal cores resembling the core of the Earth and basaltic meteorites from the 
surfaces of asteroids that had an active volcanic activity more than 4 billion years ago.
Studies of meteorites 
provide detailed information about the chronological, geochemical and 
geological evolution of the early Solar System. But unlike geological samples
from the Earth, meteorites are delivered without any information about the 
setting of the sampling site. Small asteroids,
which represents fragments of asteroid collisions in the recent past,
probably have fresh surfaces 
with minimal regolith cover and with minimal exposure to cosmic rays,
hence with a surface 
that is more representative of the interior. Fragments in the form of meteorites may therefore
more easily be linked to small asteroids than large asteroids with highly evolved surface properties.

The small asteroids are therefore vital for our understanding of mass transportation
in the inner Solar System, as well as for providing a firm basis for the dynamical
and physical relation between meteorites, NEAs and the asteroid main belt.
For a thorough discussion on asteroid research see the book by Bott\-ke et al.\ [2].

The Bering mission will consist of two fully autonomous spacecrafts which detects the asteroids, 
determine their orbital parameters, light curve and spectral characteristics.
The two spacecrafts will be identical and fly in a loose formation. 
The spacing between the two probes make it possible to 
determine the orbital parameters of the asteroid. Each probe will be able to
provide
autonomous detection, tracking, mapping and ephemeris estimation of asteroids. 
The autonomous instrumentation also include automatic linkup with Earth and
inter spacecraft communication. The autonomous operations of the
instruments are centered on the Advanced Stellar Compass [3].
 
The Bering mission will consist of two fully autonomous
spacecrafts which detects the asteroids, determine their
orbital parameters, light curve and spectral characteristics. 
A laser ranger will be used to keep track of the relative positions of the two spacecraft. Simultaneous observations from both spacecraft will allow us to accurately determine the distance to detected objects and thus make it possible to determine the orbital parameters of objects that are quickly passing out of view.
The autonomous instrumentation also include
automatic linkup with Earth and inter spacecraft communication.
The autonomous operations of the instruments
are centered on the Advanced Stellar Compass, cf. [3], [4] and [5].

 \vspace*{-0.3 cm}

\section{Asteroids and meteorites}

 \vspace*{-0.4 cm}
We have very little information on the abundance and characteristics of 
objects smaller than about 1 km except for those that have been observed in 
the immediate vicinity of Earth. The power law distribution of asteroid 
sizes suggests that objects smaller than 1 km are very abundant, see Fig.\,2. 

\begin{figure}[]
\centering
\epsfig{file=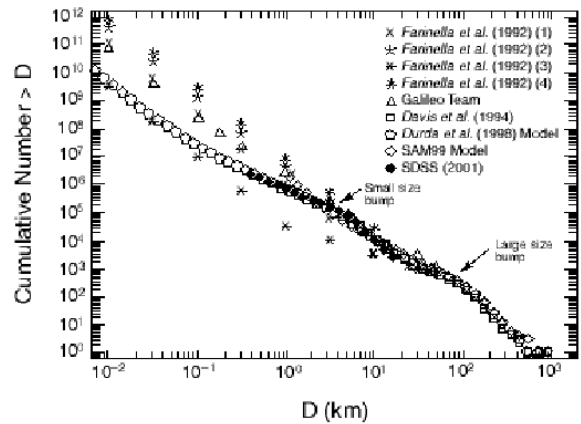,width=1.0\linewidth}
\caption{Nine different estimates of the main-belt asteroid size distribution. The small size distribution is obtained by extrapolating the observed large size trends. Figure taken from [6].
}
\label{fig:durdas}
\end{figure}

Within the asteroid belt we have no information 
about these objects, since they cannot be observed from Earth and no 
spacecraft have been actively looking for them.

The orbits of these small objects can be perturbed by physical processes
in the asteroid main belt, such as collisions.
Since any fragmentation process tends to generate power law size 
distributions of the fragments we should not be surprised to see that the 
size distribution of asteroids follow a power law distribution. There is, 
however, reason to believe that the very smallest asteroids are less abundant 
in the main belt
than a simple extrapolation of the data from the larger asteroids would suggest. 
Smaller asteroids are more easily influenced by the Yarkovsky 
effect\footnote{The effect of its rotation on the path of a small object
orbiting the Sun. Rotation causes a temperature variation, so thermal energy
is re-radiated anisotropically.} 
[7] and may thus be removed from the belt 
on a shorter time scale than the larger asteroids. Still smaller fragments 
may be removed as a consequence of the Poynting-Robertson effect\footnote{The 
effect of solar radiation on small objects orbiting the Sun, 
which causes them to spiral
slowly in. The object absorb solar energy that is streaming out radially, but
re-radiate energy equally in all directions. As a consequence there is a 
reduction in the kinetic energy, and thus in orbital velocity, which has the
effect of reducing the size of the orbit.}. A direct 
measurement of the size distribution would allow us to gain evidence of these
physical mechanisms, and how they may have influenced the development of the main
belt.

Meteorites are fragments of approximately 150 different main belt asteroids.
Since meteorites are well studied representatives of the abundant low mass 
tail of the objects that impact on Earth, a better understanding of their 
origin in the asteroid belt and their subsequent orbital evolution will 
allow us to better understand the transfer of objects from the main belt
to the NEA population.  Detailed studies of 
meteorites allow us to determine age constraints for the disruptions of 
their parent asteroids and the subsequent transfer of fragments to the 
inner Solar System. 

Asteroid photometry shows that asteroids are very diverse in terms of surface mineralogy.
The variation is equivalent to the variation observed among meteorites although a near exact match 
between the spectrum of an asteroid and a group of meteorites is extremely difficult to find. Possible reasons
for the differences between reflectance spectra of asteroids and meteorites include fine grained dust cover 
on asteroids, micro meteorite impacts on asteroids and exposure to cosmic radiation. Changes of the asteroids 
reflectance spectrum due to these poorly characterized processes are referred to as space weathering.
Only in one case has it been possible to establish a 
reasonably good case for a specific asteroid-meteorite relationship. 
The unique spectrum of the basaltic surface of 4 Vesta makes it the prime 
candidate for about 400 igneous meteorites known as the HED meteorites. 
There is considerable interest in establishing links between other groups 
of meteorites and their parent asteroids.

Unlike the larger asteroids studied from space so far, small objects are 
expected to have young surfaces, and their surfaces are therefore 
representative of their interior. A longstanding debate has been the 
relationship between the silicaceous (S-type) asteroids and the ordinary 
chondrite meteorites. Differences in spectral characteristics have been 
attributed to a poorly constrained space weathering process. Since small 
asteroids probably have smaller life times and less gravity we should 
expect them to have younger surfaces that have been exposed to space 
weathering for a shorter period of time. Also the lower gravity should 
reduce the build-up of a regolith cover on their surfaces that may hide 
geologic units underneath. Both of these effects will make a comparison 
with spectral characteristics of meteorites and other materials easier.
Data on the orbital 
distribution of objects with spectral characteristics similar to a group 
of meteorites may provide new constraints on the meteorite-asteroid 
relationship.

 \vspace*{-0.3 cm}

\section{Detection of asteroids}

 \vspace*{-0.4 cm}

\begin{figure}[]
\centering
\epsfig{file=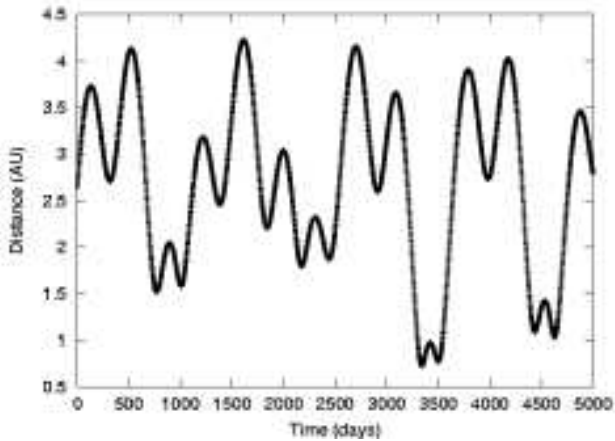,width=1.0\linewidth}
\caption{Variations in the distance to Earth as a function of time 
of the Near-Earth Asteroid 2002 NY31. Epoch of the figure is June 10, 2003 = JD 2452800.5.}
\label{fig:2002NY31distance}
\end{figure}

\begin{figure}[]
\centering
\epsfig{file=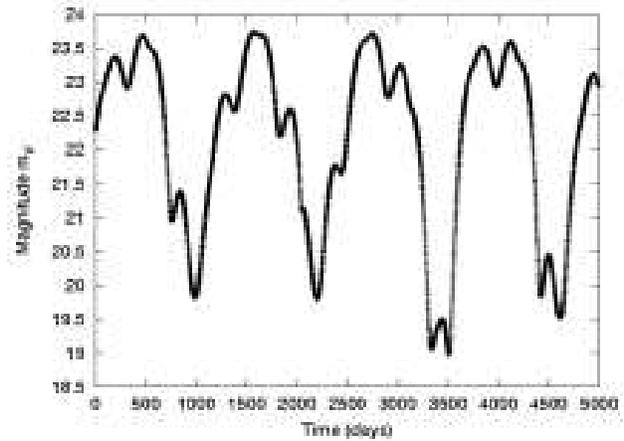,width=1.0\linewidth}
\caption{Variations in the $V$-magnitude 
($m_{V}$) as a function of time of the 
Near-Earth Asteroid 2002 
NY31 as seen from Earth. Epoch of the figure is June 10, 2003 = JD 2452800.5. 
The absolute magnitude of this object is $H=17.3$, corresponding to a diameter of around 1 km.}
\label{fig:2002NY31visibility}
\end{figure}

A number of things distinguish the brightness of an
asteroid from that of a distant star. Where the emitted radiation
from a star is due to internal nuclear processes, the brightness of
an asteroid entirely depends on reflected sunlight in terms of 
the illuminated area, as well as the albedo. This implies a 
dependency on the distances asteroid-observer and asteroid-Sun
as well as the phase angle. In total 5 parameters are needed to
describe the brightness	variations, of which 3 parameters have an 
explicit time dependency. Thus, even for a constant 
distance between
an observer and the asteroid, the brightness will vary due to 
the changing distance to the Sun, quite a different situation from 
observing remote
stars. In addition, asteroids are objects moving with velocities of the
same order of magnitude as the Earth, with distances to the Sun at the 
same order of magnitude as the distance Sun-Earth. This introduces 
brightness variations, which are not present for distant
self-luminous objects. Examples of these variations are illustrated
in Figs.\,\ref{fig:2002NY31distance} and \ref{fig:2002NY31visibility}.
 
Due to the eccentricity of $e=0.55$ of 2002 NY31, the distance
to the Earth shows a very large variation over time, in terms
of repeated periodic variations. This behavior
directly influence the $V$-magnitude ($m_V$) as seen from Earth. 
It is seen (Fig.\,\ref{fig:2002NY31visibility}) that the brightness
peaks seems to fall out from a faint background magnitude, and
that only within constrained regions is the object observable from
a telescope with a given magnitude limit. 

\begin{figure}[]
\centering
\epsfig{file=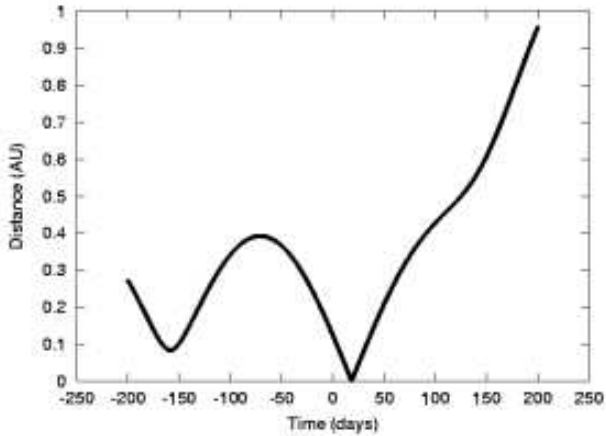,width=1.0\linewidth}
\caption{Distance to Earth of a synthetic NEA over 400 
days.
The resolution is 0.02 days = 28.8 minutes.
The figure was obtained
using modifications to the SWIFT integrator [8], see also [9] for more details. 
}
\label{fig:tpdistance}
\end{figure}

\begin{figure}[b]
\centering
\epsfig{file=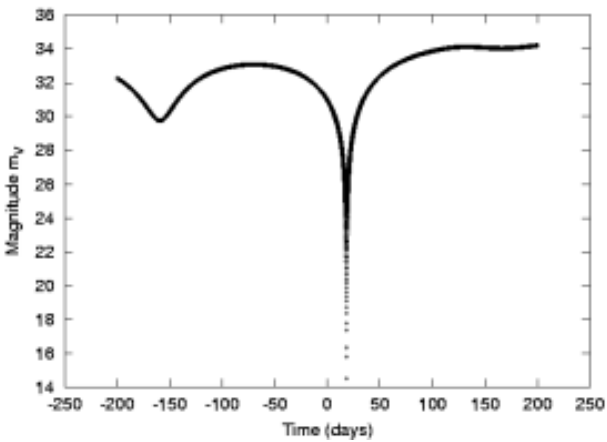,width=1.0\linewidth}
\caption{Variations in the $V$-magnitude ($m_{V}$) of a synthetic NEA as seen from
Earth. See Fig.\,\ref{fig:tpdistance} and text for details.
}
\label{fig:tpvisibility}
\end{figure}

These variations are dependent on the mentioned physical and
geometric parameters, so the smaller
the objects, the more restricted are the favorable periods of
observability. An 
example\footnote{The synthetic object has the orbital parameters 
$a=1.00583363$, $e=0.04361874$, $i=23.00978088$,
$\Omega = 29.24312401$, $\omega = 89.19306946$, $M=67.975616464$,
$epoch=JD2452000.5$, $H=32.70$, slope parameter $G=0.32$.}
of this can be found in 
Figs.\,\ref {fig:tpdistance} and \ref {fig:tpvisibility}

The synthetic object in these figures has an absolute magnitude of
$H=32.7$, corresponding to a diameter of around 1 m. It is
seen, that upon a close approach to the Earth, the magnitude
decreases drastically, from a background level above $m_V=30$
to a sudden brightness of around $m_V=15$. It is also noticed
that the brightness peak is very sharp, in fact the object magnitude
is below $m_V=20$ for merely 4.8 hours. For a telescope
with a given magnitude limit, these objects are only 
observable during the occurrence of such a brightness peak,
unless the telescope is able to reach very faint magnitudes
i.e.\ $m_V=30$.
In addition,
the brightness peak must appear while the object is within the 
field of view of the observer.
In practice, a survey telescope must constantly
monitor the whole sky in order not to miss the object due to the 
short time of visibility.

An obstacle, compared to the traditional way of making ground
based observations, is that during one night, 
each field of sky will typically only be imaged one time as
a series of three or more short exposures, and the reduction of
the images will be done during the following day or days. Due to the 
transient nature of the brightness peak of the small asteroids,
the image reduction would however have to be done real time, in order to
detect the object immediately, and initiate follow-up observations
for a verification. A ground based fast-response survey has
been proposed [10], however with a per-night
data reduction it is still remote from a real time solution.

There is an additional obstacle that must be overcome.
For sub-meter objects, the brightness peaks only occurs
upon very close approach to the observer. This means, that the
angular velocity relative to the observer becomes very large,
for the shown example it is of the order of a few hundred "/sec.
A ground based observer, using a survey telescope, will typically
need to make an optimization of the exposure time involving the
pixel size, the seeing, the astrometric accuracy and the angular 
velocity of the objects of interest.
Typically, a survey
using a large telescope going to faint magnitudes, 
will make use of a series of 60\,s--120\,s exposures.
Due to the large radial velocity, the signal
from the object will be smeared out on the detector (trailing loss). In fact, in
order to reach a resolution around 1" for the fast-moving
objects, a typical survey telescope
would be restricted to exposure times around a hundredth of a second,
presumably posing heavy demands regarding the size of the telescope.

Outside the brightness peak, the object is moving with an angular
velocity of a few "/min, so even for a telescope able to reach
$m_V=30$, the exposure time would be constrained to 60--120\,s.

In summary, a survey telescope trying to detect these small
asteroids must be capable of an
all-sky monitoring, able to perform real time data analysis, and
able to handle fast moving objects. The alternative would
be a telescope able to reach $m_V=30$, or beyond, with exposure
times per image frame limited to a few minutes. For the latter 
option, our conjecture is that such a telescope is currently not
technologically feasible.  For the first option, we propose
Bering as a solution to meet these requirements in terms
of the autonomy and the application of the Advanced Stellar Compass.
In addition to the simple object detection, discussed in the above,
Bering also provides the possibility of performing 
spectroscopy/photometry of the feasible objects, thus adding another
dimension of performance in comparison to ground based observations.

 \vspace*{-0.3 cm}

\section{Mission profile requirements}

 \vspace*{-0.4 cm}

For the reasons outlined above, the main goal of the proposed Bering mission 
is to detect and characterize a sizable amount of sub-kilometer objects 
from space. This will be the first systematic survey of sub-kilometer objects 
in the Solar System. The numbers detected need to be sufficiently high that 
the distribution of small objects with similar spectral characteristics and 
therefore potentially identical parent asteroid may be established throughout 
the main belt. 

Due to the transient visibility variations,
the probes 
must detect the objects and guide the science instruments in a fully 
automated process. 

The two spacecrafts will each carry an Advanced Stellar Compass (ASC) system
with 7 camera heads, a folding mirror based multi-spectral telescope imager,
magnetometers and a laser-ranger. These instruments will make it possible to
obtain orbital parameters, light curves, rotation state, surface composition, and in
some cases albedo, size, high quality images, mass and magnetic properties, cf.\,[12].

\begin{figure}
\centering
\epsfig{file=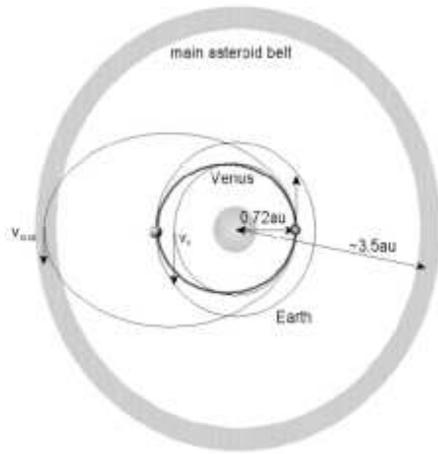,width=1.0\linewidth}
\caption{Mission profile. Planet and asteroid orbits are shown circular and
coplanar for simplicity. The figure also show the positions of Earth and Venus
during the first phase of the mission and it assumes a fly-by at the second
passage of the perihelion (1.5 period).}
\label{fig:orbit}
\end{figure}

In order to determine the distribution and dynamics of small objects and 
their links to the NEA po\-pu\-lation we need to detect 
objects in the main 
asteroid belt as well as inside the Earth’s orbit. The objects within the 
mean motion
resonances in the asteroid belt
 that are either already NEAs or are becoming NEAs within a short 
time frame, generally have aphelion within the main belt and spend most of 
their time outside the Earth orbit. 
We therefore propose a mission profile that would give 
us data on the distribution of small objects all the way from 0.7 AU to 3.5 AUs,
see Fig.\,\ref{fig:orbit}. Such an orbit can be achieved with a single
unpowered gravity assist maneuver from Venus whereas the $\Delta$V required to 
reach Venus could be delivered directly by the launcher, which could be of the 
Soyuz class.

For each detected object we propose to automatically determine: \\
- The heliocentric position, and velocity vector of the \\ \hspace*{0.3 cm} object. \\
- Multicolor photometry of the reflected light from the \\ \hspace*{0.3 cm} object in the 
350–-2200 nm range. \\
- The light curve of the object and thereby its rotation \\ \hspace*{0.3 cm} period.

For a few selected larger objects we furthermore propose to:\\
- Record multicolor images of the surface. \\
- Determine the mass and magnetic moment of the object. \\

The data will allow us to determine a current orbit for the object. With a 
detection at $m_{V} =9$ for the Advanced Stellar Compass and a detection 
limit of $m_V = 25$ for the multicolor imager we will be able to follow the 
object out to a distance of approximately 1500 times the distance where 
it was detected. Depending on the geometry and the size of the object this 
will typically allow us to follow the object for days to weeks and 
determine high precision orbital parameters for a large fraction of the 
orbital arc. The orbital data will also allow us to determine the objects 
position in orbital space and its proximity to resonance’s and/or other 
asteroids or asteroid families with similar spectral characteristics 
and dy\-na\-mical characteristics.

The photometry will allow us to determine the spectral type of the object. 
This will make it possible for us to determine its relationship with other 
asteroids and/or groups of meteorites with similar mineralogy. Ultimately, 
we will attempt to backtrack the object to its parent asteroid

The ability to detect objects down to $m_{V} = 25$ from within the asteroid belt 
with the multicolor imager may also be utilized to further constrain the 
density of small objects. In a few campaigns we propose to make a series of 
frames with either subsequent on board processing or data processing on 
Earth. This will allow us to determine the number of asteroidal objects in 
each frame, going to very faint objects. Although the size-distance 
relationship cannot be determined on the basis of a few frames these data 
may be used to check predictions based on models of the distribution of 
asteroids.

 \vspace*{-0.3 cm}

\section{Object detection rate}

 \vspace*{-0.4 cm}
The requirement to the number of objects Bering would detect per 
unit of time is important to quantize both due to the
scientific feasibility estimate, as well as to formulate the design requirements
of Bering.
Due to the complex variations in brightness, this is not trivial
to estimate. 

For an observer moving in an inertial frame,
it would maybe be possible to estimate the flux of objects through
the detection sphere of Bering. However, due to the proper motion
of Bering, and due to the strong dependency on geometry, it 
becomes a more complex task to evaluate the flux of objects.
A possible approach could be to apply the estimates of 
collision rates in the asteroid belt [11], however
for a detailed examination of the Bering mission profile, we need
a more direct approach [9].

\begin{figure}
\centering
\epsfig{file=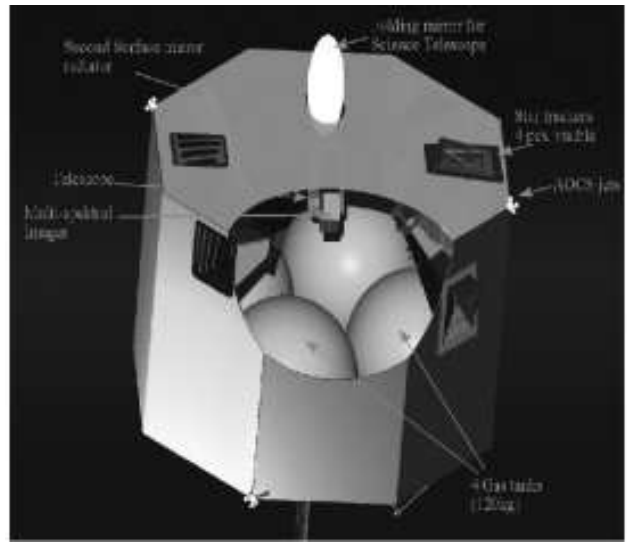,width=1.0\linewidth}
\caption{Artistic impression of the Bering preliminary design.
}
\label{spacecraft}
\end{figure}

 \vspace*{-0.3 cm}
\section{Spacecraft design}

 \vspace*{-0.4 cm}
The Bering spacecrafts have to possess a high degree of autonomy since at a
distance of 3 AU the closed loop to Earth is about 48 min. The autonomy must
encompass all everyday routines, such as object detection, classification,
tracking and initial data acquisition. 
An overview of the payload is provided in Table \ref{tab:payload}.

The core autonomy of the proposed mission is centered on the Autonomous 
Stellar Compass (ASC), a fully autonomously operation star tracker [3].  The 
ASC consist of a powerful microcomputer with search engine and star 
catalog, with several camera heads attached, typically two to four. Each 
camera head will acquire an image of the night sky in the Field Of View (FOV) 
two times per second and pass on the digital image to the data processor, see [5].

Since all luminous objects above the selected threshold are detected, 
non stellar such as galaxies, nebulae, other satellites and asteroids are 
also autonomously picked out. They are identified as luminous objects brighter 
than the threshold, with no matching object in the star catalog. Since 
the ASC has established the attitude of the image in question, the apparent 
position of the object is established with high accuracy. Typically the 
instant position determination is in the range of 3 arc seconds, but since 
the ASC update the measurement per camera twice per second, averaging will 
bring the accuracy to the sub-arc second level in a matter of seconds. In 
this way, the entire FOV is searched for non-stellar objects. Since the FOV 
covers 1\% of the entire night sky, a scanning operation of the spacecraft 
will have to be employed. We plan to arrange 7 camera heads on board the 
spacecraft, whereby almost 100\% sky coverage is guaranteed at a spin rate 
of once per hour.

The non-stellar objects are entered into a database. When the same area of 
the night sky is revisited after one rotation period of the spacecraft, the 
apparent position of the objects are established again. Any object thus 
showing a proper motion above the measurement noise is moved to an asteroid 
candidate list for further investigations [13].

The main science instrument on the proposed mission is a multi-band imaging 
telescope, see Fig.\,\ref{spacecraft} and [5]. The telescope might have a focal length of 1~m, 
and an entrance pupil of about 0.3~m. To compensate for the spacecraft rotation 
and to allow for fast tracking, the telescope is fitted with a folding 
mirror. To enable fast and accurate tracking of a target, the telescope is 
furthermore equipped with an ASC camera, mounted on the back of the 
secondary mirror. The ASC camera and the telescope optical axes are 
approximately parallel. From simultaneous images of the night sky, the 
relative orientation of the telescope to the ASC camera is easily 
established with high accuracy.

\begin{table}
\caption{Scientific payload.}  
\begin{tabular}{|l|l|l|}  \hline
& & \\ 
{\bf Instrument} & {\bf Science} & {\bf Mission}  \\ 
  & & {\bf support} \\ 
 & & \\ \hline \hline
& & \\
Advanced  & Detection of & Additude and    \\
Stellar  & asteroids  & navigation  \\
Compass &   &  \\
 & & \\
Multiband & Surface composition, & Location \\
Imaging &  rotation and spatial & of sister  \\
System & extend of asteroid, & spacecraft \\
& accurate apparent &  \\
 & position of target &  \\
 & & \\
Laser  & Exact distance to & Ranging of  \\ 
Ranger   & close asteroids & sister space- \\
& & craft  \\
& & \\
Magnetometer & Magnetic properties, & Solar \\
Probes & gravity estimate of & pressure \\
       & asteroid and thereby & estimation  \\
  &      mass and density   & \\  
& & \\ \hline
\end{tabular}
\label{tab:payload}
\end{table}

When the ASC has detected a Sun orbiting body, the apparent inertial position 
of the body is measured and passed on to the main onboard processor, that at
a convenient time, points the telescope at the target. This pointing is greatly
facilitated by the ASC camera that is co-aligned with the telescope,cf.\,[12] and [5].  The multi-spectral 
imager serves multiple functions too. Its main objective is to characterize
the target through spectral analysis and to obtain high fidelity multispectral
science images. However, through the guide information from the ASC, the stars in
the telescope image are analyzed, whereby an extremely accurate  apparent position of
the target is found.
This procedure allows the Bering spacecrafts to keep track of more than 1000 objects 
at any give time of the mission.

The laser ranger will determine the distance to all objects that happen to pass
close by the spacecraft. Small free-flying magnetometer probes will be launched toward
selected larger objects that comes within close range of the spacecrafts. At close
encounter the magnetometers will provide information about the distribution of the
magnetic field, and as such probe the interior of the asteroid. The trajectory of the magnetometer
probe will give information about the mass and thereby the density of the asteroid.

 \vspace*{-0.3 cm}
\section{Summary}

 \vspace*{-0.4 cm}
Bering is a deep space mission to detect and characterize sub-kilometer 
objects between Jupiter and Venus. The focus on the mission is on the investigation 
of asteroid evolution, transfer of asteroids from the main belt to the inner Solar System,
 and determination of meteoritic parenthood. The
spacecrafts will carry advanced stellar compasses, a multi spectral imager,
a laser ranger and magnetometer probes.

\end{document}